# Towards a Quantitative, Metabolic Theory for Mammalian Sleep

Van M. Savage*¶★ and Geoffrey B. West*¶

*Santa Fe Institute, 1399 Hyde Park Rd., Santa Fe, NM 87501
¶Los Alamos National Laboratory, MSB285, Los Alamos, NM 87545
★Bauer Center for Genomics Research, Harvard University, 7 Divinity Ave., Cambridge, MA 02138

# Abstract

Sleep is one of the most noticeable and widespread phenomena occurring in multicellular animals(1). Nevertheless, no consensus for a theory of its origins has emerged. In particular, no explicit, *quantitative* theory exists that elucidates or distinguishes between the myriad hypotheses proposed for sleep. Here, we develop a general, quantitative theory for mammalian sleep that relates many of its fundamental parameters to metabolic rate and body size. Most mechanisms suggested for the function of sleep can be placed in this framework, e.g., cellular repair of damage caused by metabolic processes and cortical reorganization to process sensory input. Our theory leads to predictions for sleep time, sleep cycle time, and REM (rapid-eye-movement) time as functions of body and brain mass, and explains, for example, why mice sleep ~ 14 hours per day relative to the 3.5 hours per day that elephants sleep. Data for 96 species of mammals, spanning six orders of magnitude in body size, are consistent with these predictions and provide strong evidence that time scales for sleep are set by the brain's, not the whole-body, metabolic rate.

In contrast to other obvious physiological phenomena such as eating, breathing, and walking, neither the function nor the mechanism by which sleep occurs is well understood. Indeed, the quest for a fundamental theory of sleep is considered one of the most important, unsolved problems in science (2-6). Recent neurobiological studies have made great advances in understanding the mechanisms involved in sleep. Hormones, cells, and enzymes whose levels of activity and expression vary during sleep and between sleeping and waking states have been identified (2, 3, 7-9). Although these studies have been unable to determine the purpose of sleep, such investigations will play an increasingly important role in determining what processes are specific to sleep.

Among the most studied and best known hypotheses for the purpose of sleep are: (*i*) rest for the body or brain (11-13); (*ii*) cortical reorganization and processing associated with memory and learning (14-17); and (*iii*) cellular repair in the body or brain (3, 18-20). Here, we develop a quantitative theory whose general structure can, in principle, be used to test any of the above hypotheses through analyses of inter-specific data with the ultimate goal of distinguishing between them. Our starting point is the observation that these suggested underlying processes are related to metabolic rate. We speculate a) that unlike the brain, other organs or tissue can be repaired or reorganized on a continuous basis during waking or resting periods without significantly interfering with "normal" activity and functionality(5), and thus, that the brain is exceptional in comparison to other organs and tissues because it requires a special state, identified as sleep, devoted primarily to these critical activities; b) that periods of cortical reorganization and/or repair of neuronal damage are primarily induced by activity during wakefulness (AW). Stated slightly differently, we posit that the foremost function of sleep is to counteract *secondary*, detrimental effects of metabolism and is not directly connected to the *primary* effects of metabolism--providing the necessary energy and nutrients for normal functioning. This immediately suggests that sleep times might scale differently, even inversely, to typical biological times, and indeed, that is what we predict theoretically and find empirically in our analyses below. Cellular repair is especially crucial for the brain because, unlike most other organs and tissues whose cells are continually being replaced, neurons are typically not regenerated during the lifetime of the organism. This constancy of neurons suggests that most changes occurring in the brain are due to reorganization of connections between existing neurons. Furthermore, faithful repair of cellular damage is critical for maintaining the long-term integrity and functioning of the brain.

Dissipative energy is necessarily produced as a by-product (secondary effect) of metabolic processes and inevitably leads to cellular damage. Dissipation occurs both at the basic biochemical level as, for example, in the production of free radicals, and at the microcapillary level in the circulatory system due to hemodynamic analogs of viscous drag forces. Experimental evidence that brain cell damage is caused by sleep deprivation (21, 22) has been reported, although other studies did not corroborate these findings (23, 24). Furthermore, the recent findings of Cirelli et al. demonstrate that *Drosophila* Shaker mutants with reduced sleep time experience a concurrent and identical decrease in lifespan(10). Assuming all other factors such as metabolic rate are held constant-- consistent with measures of activity(10)--these reduced sleep times result in increased periods of damage and reduced periods of repair. Thus, lifespan for these Drosophila Shaker mutants should be reduced, just as observed, further suggesting sleep is linked to the repair of damage(10). In addition, there have been several studies that show sleep time is positively correlated to metabolic rate, body weight, brain weight, and other physiological variables (11, 13, 25). Since cellular damage is directly tied to metabolism, and its repair is a leading hypothesis for sleep, we develop our theory with this scenario in mind. However, the framework's general structure can accommodate any mechanism that links sleep to metabolic rate.

Smaller animals have proportionately higher metabolic rates per unit mass than larger animals, so, according to these ideas, they require more sleep over a fixed period of time. Such qualitative predictions that agree with trends in the data are clearly useful, but much more detailed analyses and tests of hypotheses are possible once a quantitative theory has been developed. Below, explicit relationships between various facets of sleep, metabolic rate, and body size are derived. These provide a way to distinguish whether cellular repair during sleep occurs solely in the brain or throughout the entire body, and whether cellular damage occurs at the same rate during sleep and wakefulness.

**Theory**

**Mass dependence of ratio of sleep time to awake time.** We begin by assuming that most cellular damage (secondary effect) is caused by metabolic processes (primary effects) during wakefulness and that cellular repair (response to secondary effect) and/or reorganization primarily occurs during sleep. These assumptions will be relaxed below. Dissipative metabolic energy, e.g. viscous-type forces in metabolite transportation networks, including the exchange of oxygen from hemoglobin to tissue, occurs primarily in network terminal units (capillaries and mitochondria), which are approximately constant in size and do not vary appreciably with body mass (26). This invariance implies that both the total rate of dissipative metabolic energy production and the metabolic rate itself, $B$, scale linearly with the number of such terminal units. Consequently, the total rate of dissipative metabolic energy production causing cellular damage is a mass-independent fraction, $f$, of metabolic rate, $B$. This invariant fraction is calculable at the capillary level(26). At the mitochondrial level, it also holds because many of the core metabolic reactions are shared across mammals, each yielding a set amount of energy with free radicals as byproducts.

The rate at which dissipative metabolic energy is deposited on and in an average cell is therefore $fB_c$, where $B_c$ is the average *in vivo* cellular metabolic rate. The total metabolic rate of the organ, tissue sample, or whole body being considered is $B = N_c B_c$, where $N_c$ is the corresponding total number of cells. During sleep the total amount of energy causing damage (secondary effect) that needs to be repaired is therefore given by $fBt_A$, where $t_A$ is total AW time per day. We assume that cellular repair is a relatively uniform, local phenomenon that is common to mammals and occurs at a level significantly smaller than the cell itself so that the power density (per unit volume) required for repair, $P_R$, is independent of body size (see Appendix for a more detailed discussion). For faithful repair, energy spent in damage while awake must be compensated for by energy devoted to repair during sleep: $fBt_A = \varepsilon P_R v_c N_c t_S$, where $v_c$ is the average cell volume, $t_S$ is sleep time per day, and $\varepsilon$ is the efficiency of repair—the ratio of energy given to damage relative to the energy required to repair that damage. The ratio $\varepsilon$ is dictated by thermodynamics and cellular structure and is assumed to be independent of body size. (Note that our conclusions hold so long as $\varepsilon \propto P_R^{-1}$). Since total volume considered is $V = N_c v_c$, we have

$$\frac{t_s}{t_A} \approx \left(\frac{\rho f}{\varepsilon P_R}\right)\frac{B}{M} \qquad (1)$$

where $M$ is the corresponding mass, and $\rho \equiv M/V$, the tissue density, which is constant. Eq. (1) can be expressed in terms of purely cellular quantities as $t_S/t_A \approx (\rho f/\varepsilon P_R)B_c/m_c$, where $m_c$ is the mass of the average cell. If cortical reorganization is the origin of sleep, Eq. (1) still holds but with $P_R$ now interpreted as the power density required for affecting such an activity, $f$ the fraction of metabolic energy used for processing information or sensory input, and $\varepsilon$ the measure of efficiency in converting sensory input to cortical organization. In either case, the quantity related to metabolic rate is neither sleep nor awake time but rather, a ratio of the two.

The scaling of whole-body metabolic rate with body size, $B = B_0 M^{3/4}$, where $B_0$ is the normalization constant, has been shown to reflect general properties of resource distribution networks. These networks are assumed to be space filling and have invariant terminal units. Minimizing energy expenditure in the circulatory system leads to quarter-power allometric scaling for many physiological rates and times (26). Organs such as the brain, which are supplied by major arteries, behave as nearly autonomous subunits and can effectively be treated as independent systems subject to the constraints of the theory. Their metabolic rate, $B_i$, is therefore predicted to scale approximately as $M_i^{3/4}$, where $M_i$ is the organ mass. If $M_i \propto M^{a_i}$, then $B_i \propto M^{(3/4)a_i}$ and $B_i/M_i \propto M_i^{-1/4} \propto M^{-p_i}$ with $p_i \equiv (1/4)a_i$. For most organs the allometric exponent $a_i \approx 1$, so $B_i \propto M_i^{3/4}$. Brains, however, are exceptional. Reported empirical values for $a_b$ ($b$ denotes brain) range from 0.65 ($\approx 2/3$) to 0.76 ($\approx 3/4$) (27-29). Consequently, $B_b/M_b \propto M_b^{-1/4} \propto M^{-p_b}$ with $p_b$ ranging from about 1/6 to 3/16. Unfortunately, no *in vivo* data are available to test this.

However, the limited available data for rates of oxygen consumption in brain tissue *in vitro* are not inconsistent with these predictions (see Appendix).

Thus, if the primary purpose of sleep is to repair damage done to brain cells, then

$$\frac{t_S}{t_A} \propto M_b^{-1/4} \propto M^{-p_b} \qquad (2)$$

with $p_b$ in the range $0.16 - 0.19$. Note that the *absolute* value of $t_S/t_A$ could be predicted from Eq. (1) if repair and dissipative rates ($P_R$ and $f$) were known. If the primary purpose were to repair damage to organs other than the brain, the exponent in Eq. (2) would be -1/4 rather than -1/6 or -3/16. This provides a way for distinguishing whether repair during sleep occurs in cells throughout the body or primarily in the brain.

**Extension to include damage during sleep.** Eqs. (1) and (2) can be further generalized to include damage occurring during sleep itself:

$$\frac{t_A}{t_S} \approx \frac{\varepsilon P_R M}{\rho (fB)_A} - \alpha \approx \frac{\varepsilon P_R M^{p_b}}{\rho (fB_0)_A} - \alpha \qquad (3)$$

where $\alpha \equiv (fB)_S / (fB)_A$ is the damage rate during sleep relative to that during AW.

If power devoted to repair far exceeds that leading to damage during sleep, i.e., $\varepsilon P_R M / \rho \gg (fB)_S = \alpha (fB)_A$, the additional contribution, $-\alpha$, can be ignored. Since $t_A/t_S > 0$, it follows that $\varepsilon P_R M_b^{1/4} / \rho > \alpha (fB_0)_A$, so the contribution of $\alpha$ decreases with increasing $M_b$ (or $M$), only being important for small mammals. For example, if $\alpha \sim 1/3$, as suggested by data (see Appendix), then damage during sleep is significant only for very small animals that sleep more than ~18 hours/day.

If damage rates were unchanged between sleeping and waking states ($\alpha \approx 1$), then, since $t_S + t_A = 1$ day, Eq. (3) reduces to a pure power law for $t_S$: $t_S \propto M^{-p_b}$. This is inconsistent with fits to data (see below), and with the previously mentioned value of $\alpha \sim 1/3$. It is known that whole-body metabolic rate only decreases by 10-15% in sleep versus resting states, and the brain's metabolic rate decreases by much less (4). Thus, for damage rates to decrease significantly during sleep, $f$ must decrease, presumably by mechanisms that, for example, suppress radical production or increase anti-oxidants. Regardless of mechanism, Eq. (3) provides a method for probing this by determining $\alpha$.

This analysis suggests that the primary quantity to consider is the ratio $t_S/t_A$ rather than either $t_S$ or $t_A$ separately, as has been typically done in the literature. Most physiological rates (such as heart and respiratory rates) scale as mass-specific metabolic rate, $B/M \propto M^{-1/4}$, whereas most physiological times (such as blood circulation time and time to maturity) scale as its inverse, $M^{1/4}$. These scaling behaviors originate from generic constraints on distribution networks (26, 30) and pertain to quantities that are linked to the *primary* beneficiary purpose of metabolism: to supply energy and nutrients. This is in distinct contrast to sleep time, $t_S$, whose purpose is to counteract detrimental, *secondary* effects of metabolic processes. This critical difference is reflected in the scaling of $t_S$, which not only does not follow a simple power law, but, unlike almost all physiological times, *decreases*, rather than increases, with body size.

**Mass dependence of sleep cycle time and fraction of sleep time spent in REM.** Two of the most intriguing aspects of sleep are the division between REM and non-REM (NREM) states and the oscillations between them known as the sleep cycle. Understanding these phenomena could help reveal the role that each plays in brain repair and functioning. If, unlike sleep itself, they are directly driven by the primary beneficiary purpose of metabolism, all times associated with them are expected to scale like typical physiological times. As such, sleep cycle time, $t_c$, the time between the endings of periods of REM sleep, is predicted to scale as $t_c \propto M_b^{1/4} \propto M^{P_b}$. Consequently, if $n_C$ is the number of sleep cycles per day, so total sleep time $t_S = n_C t_C$, then $n_C (\propto t_S / t_C)$ decreases with $M$ but does not follow a simple power law. The average length of both REM sleep per cycle, $t_R$, and of NREM sleep per cycle, $t_{NR}$, are likewise predicted to scale as $t_R \propto t_{NR} \propto t_C \propto M_b^{1/4} \propto M^{P_b}$. So, during a time interval that spans several sleep cycles, the fraction of sleep time spent in REM, $R = n_C t_R / t_S = t_R / t_C$, is predicted to be independent of mass. A possible mechanism for the division between REM and NREM states is that they represent distinct periods devoted primarily to reorganizing or repairing different functional components or cell types or tissue within the brain. Local regions that become activated during REM cycles have been identified and knowledge of their size would allow an estimate for $R$.

**Methods**

All of the sleep data are from Zepelin(31), and methods are detailed therein. Data for body mass are not given in Zepelin(31), so most values for body mass come from (17) in the text, which is an earlier compilation that Zepelin(31) draws from heavily. When sleep measurements listed in Zepelin(31) were not given in or did not match sleep data in (17) in the text, the masses given in Smith *et al.*(32) were taken, and when masses could not be found there, we used the average of the range of values given in Nowak(33). All of the data and the sources for the body mass values are listed in the supplementary data table. Some of the original sources given in Zepelin(31) were consulted to determine which species were used. In a few cases, the logarithmic averages of body masses were calculated for groups of species (*e.g.*, four species of *Microtus* and five species of

*Peromyscus*). This was done in order to be consistent with the original sleep data in Zepelin (31). Data for brain mass were taken from (17) in the text, but two values were excluded because the sleep values given did not match those in given in Zepelin(31).

Certain marine mammals sleep with one hemisphere of the brain at a time. Since in Eq. (1) the variable $t_S$ is the amount of sleep time per cell or per tissue, total sleep time for these marine mammals must be divided by two to obtain the appropriate sleep time. (There is often an asymmetry between amounts of sleep for the left and right hemisphere in marine mammals, so dividing by two should be regarded as the average sleep per neuron(34).) The original sleep data from Zepelin(31) were adjusted accordingly for the three species: *Tursiops truncates*, *Globiocephalus scammoni*, and *Phocoena phocoena*. Moreover, marine mammals are known to have very small or perhaps non-existent amounts of REM sleep, and in accord with several other studies, the data for REM sleep for *Tursiops truncates* and *Phocoena phocoena* were excluded (31). Data for the REM sleep of *Globiocephalus scammoni* was included and corresponds to the large outlier in Fig. (3). Finally, REM sleep time and sleep cycle time for *Elephas maximus* were excluded because Zepelin(31) denotes those values as doubtful, and the REM sleep time of 0.0 for *Tachyglossus aculeatus* was excluded because recent studies have shown that differentiating between REM and NREM sleep for this species is especially difficult (35).

Allometric exponents were determined using OLS regression on ln-ln plots of the data. Confidence intervals and p-values were computed using Mathematica.

## Results

**Sleep time to awake time ratio.** We now compare our predictions with data (see supplementary data). Fig. 1 shows daily values of $\ln(t_S/t_A)$ versus $\ln(M)$ for 83 taxa, representing 96 species and 79 genera of mammals and spanning six orders of magnitude in body mass. The slope, obtained by OLS (Ordinary Least Squares) regression, is –0.16 (p<0.0001, n= 83, 95% CI: –0.21, -0.11), in agreement with Eq. (2). Note that the CI include both -1/6 and -3/16 (values linked to repair in the brain) but *exclude* -1/4 (the value predicted if repair is occurring throughout the body), consistent with the hypothesis that the major purpose of sleep is repair to neuronal cells. This exponent is in clear disagreement with +1/4, the naive expectation if $t_S$ behaves like a typical physiological time. Not only does it have the wrong magnitude but, more significantly, the *wrong sign*. In Fig. 2, using data for brain mass, we plot $\ln(t_S/t_A)$ versus $\ln(M_b)$, which has a slope of -0.21 (p<0.0001, n=56, 95% CI: -0.28, -0.14), in agreement with Eq. (2) and the -1/4 predicted if the brain's metabolic rate drives sleep. Recall that $t_S$ satisfies a pure power law when damage is uniform throughout sleeping and waking periods ($\alpha \approx 1$). Data give an exponent for $t_S$ of -0.1 with CI that do *not* include the values -1/6, -3/16, or -1/4 predicted if $\alpha \approx 1$. Moreover, by fixing $p_b \approx 1/6$ or 3/16, $\alpha$ is determined to be small (see Appendix). These results are consistent with damage during sleep being negligible. Although gathering more data would improve estimates for $\alpha$, the scatter in Figs. 1 and 2 is not just measurement error but is due to inherent biological effects such as alternate life-histories or predator-prey relationships (25, 31).

**Sleep cycle time and fraction of sleep time spent in REM.** In Fig. 3 we show $\ln(t_C)$ versus $\ln(M)$ for 32 species of mammals, spanning five orders of magnitude. Using OLS regression we find a slope of 0.19 (p<0.0001, n=32, 95% CI: 0.14, 0.23), which is close to 3/16. The CI include the predicted 1/6 and 3/16 but *exclude* 1/4, thereby providing support that sleep cycle time is directly set by the brain's metabolic rate. Finally, in Fig. 4 we show ln(*R*) versus ln(*M*), for 59 species of mammals, spanning five orders of magnitude. The mean value of *R* is 0.17, and the slope in Fig. 4 is -6e-3 (p=0.74, n=61, 95% CI: -0.04, 0.03), consistent with REM sleep being a constant fraction of total sleep time, as predicted. See Elgar, Pagel, and Harvey for additional supporting evidence (13).

**Discussion**

Since lifespan scales like a typical physiological time ($\sim M^{1/4}$) (28), smaller mammals spend less time asleep on average during a lifetime than larger ones. Conversely, larger animals spend a much smaller *fraction* of their lives asleep than smaller ones. These observations raise interesting questions about the possible ecological and evolutionary consequences of sleep (e.g., predator–prey effects) and the corresponding advantages and/or disadvantages it confers on different size mammals(25, 36). Indeed, the fraction of time spent asleep must eventually become limiting for small animals because some amount of time is necessary just to forage, to feed, and to reproduce(13, 36). Including such effects could enhance this model. Another natural extension of our analyses is to mammals during ontogenetic growth and to other taxa. This is particularly interesting for birds because their brain allometry and partial use of unihemispheric sleep distinguish them from most mammals (1, 2, 28), and for ectotherms because the temperature dependence of metabolic rate as well as the variation in cell and genome size adds other parameters to the theory (see Appendix) (30). Unfortunately, in both cases there are insufficient data available for definitive analyses.

We have presented a theory to address sleep-related questions in a well-defined, general framework whose quantitative predictions can be directly compared to existing data. Our results demonstrate that there is substantial evidence that the function of sleep is related to metabolic processes in the brain and, in particular, to the repair of neuronal damage—a response to the secondary effects of metabolism (3, 18-20). The theory is able to explain sleep times that differ by a factor of seven across organisms that differ in mass by six orders of magnitude, revealing why a mouse sleeps four times more per day than an elephant. Further work needs to be done to understand the detailed mechanisms underlying damage and repair and their relationship to metabolic rate and sleep. Our work suggests experiments: 1) tests of whether the power density of cellular repair is body mass independent, 2) measurements of the fraction of metabolic energy given to damage, and 3) determination of the scaling of brain metabolic rate with body size. Our results strongly suggest that metabolic processes in the brain control the underlying mechanisms for the function of sleep and are a major determinant of why sleep cycle time increases with body size and why mammals need the amounts of sleep they do.

## Appendix

**Dependence of power density given to repair, $P_r$, on body size, $M$.** We assume that the power density (per unit volume) given to repair, $P_r$, is constant. This is because we are assuming the repair process is occurring locally, at a level significantly smaller than that of the cell, and because we assume that the entire cell, including cell walls, proteins, and the cytoskeleton, requires repair.

Depending on the exact mechanisms of cellular damage and repair, however, it may be most appropriate to focus on other quantities. For example, if DNA is the dominant site of damage and repair, it is most appropriate to consider the power given to repair per DNA content. If power given to repair per nucleotide or per codon is constant due to the limited spatial access to these domains, then it follows that power given to repair per DNA content is constant. For mammals DNA content and glial cell size are roughly invariant, but neuronal cell size appears to vary as $M_b^{1/4}$(37-41). Consequently, within this scenario, power given to repair per unit volume, $P_r$, would still be a constant for glial cells but would decrease as $M_b^{-1/4}$ for neuronal cells. Thus, if glial cells require the most repair, the theory is exactly the same as presented in the main text. However, if neuronal cells require the most repair during sleep, most repair at the cellular level is to DNA, and power given to repair per DNA content is constant, then the mass dependence of $P_r$ decreases as $M_b^{-1/4}$, so the ratio of sleep time to awake time would be predicted to be mass independent. This prediction is in clear disagreement with data and suggests that this scenario is incorrect, but it does provide a means for falsifying our theory. That is, if this mass dependence for $P_r$ was supported by empirical findings at the cellular level, it would suggest that our theory is incorrect, and the influence of metabolic rate on sleep time may need to be re-evaluated. Alternatively, if $B_b/M_b$ was constant, which contradicts our assumption and limited empirical data (see next section), and $P_r$ decreased as $M_b^{-1/4}$, the predictions of our theory would be exactly the same as presented in the main text, but the source of the mass dependence would come from the power density given to repair and *not* from the mass-specific metabolic rate of the brain. Measurements of $P_r$ and $p_b$ are therefore crucial for determining both the source of mass dependencies in our theory and for assessing the validity of our theory.

Within ectotherms cell size and genome size vary over several orders of magnitude, and this variation may be related to mass-specific metabolic rate(42, 43). Therefore, to extend this framework to ectotherms, it will be crucial to discern between the different mechanisms given above. This problem is somewhat simplified by the fact that cell size and genome size appear to vary together linearly.

Since we do not know the exact nature of the repair occurring at the cellular level (either what is being repaired or the mechanism by which it is repaired), we focus on the power per unit volume in the main text because its generality does not depend on a specific repair mechanism. As explained above, however, the framework can be modified to accommodate different and/or more complicated repair mechanisms.

***In vitro* data for scaling exponent of brain metabolic rate with body mass, $p_b$.** Qualitative support for a negative value for $p_b$ is found in Elliott and Henderson(44), Davies(45), and Tower and Young(46). Elliott and Henderson(44) measured rates of oxygen consumption in cortical tissue for three species of mammals (rat, cat, and cattle). Their measurements were taken *in vitro* within one hour of sacrifice of the animals. They concluded that $p_b \approx 0.1$. Subsequent studies by Davies(45) concluded that *in vitro* measurements are usually an underestimate of *in vivo* values, but using a limited amount of *in vitro* data for rates of oxygen consumption in brain tissue, he obtained a value of $p_b \approx 0.07$. Later, Tower and Young (46) performed the same measurements as Elliott and Henderson(44) with inclusion of data for fin and sperm whale and obtained similar results. Tower and Young also used data for seven species to show that the activity of acetylcholinesterase in the dorsal cerebral cortex decreases with brain mass as $M_b^{-0.2}$. Although these findings are based on very limited amounts of *in vitro* data (as opposed to the *in vivo* data needed to test directly this theory), the scaling exponents are not inconsistent with the theoretical values of $-1/6$ and $-3/16$.

**Estimate of fraction of metabolic rate given to damage during sleep, $\alpha$.** We performed a linear plot for $t_A/t_S$ versus $M^{-3/16}$ and versus $M^{-1/6}$. From Eq. (3) the intercept of these plots corresponds to $\alpha$. Using OLS regression, we obtained the value $\alpha=0.35$ (p=0.074, n=83, 95% CI: -0.04, 0.75) for $p_b=-3/16$ and $\alpha=0.27$ (p=0.20, n=83, 95% CI: -0.15, 0.70) for $p_b=-1/6$. The confidence intervals are large but suggest $\alpha<1$. Note that OLS regression assumes homoscedascity (uniform standard deviation for each datum), which is not strictly satisfied in these plots. The mass dependence for the standard deviation of each datum in this plot behaves approximately like $\alpha/M^{19/16}$ and $\alpha/M^{7/6}$, respectively, where $\alpha$ is the standard deviation in mass for each species. The relative standard deviation, $\alpha/M$, is roughly constant for mammals(47), so the standard deviation for each datum in this plot slightly decreases with $M$. Consequently, the errors for $\alpha$ are even larger than the ones given here, but the data likely still suggest $\alpha<1$.

VMS and GBW are grateful for the support of the Thaw Charitable Trust, a Packard Interdisciplinary Science Grant, the National Science Foundation, and Los Alamos National Laboratory. VMS also acknowledges support from the National Institutes of Health through the Bauer Center for Genomics Research. We thank Morgan Ernest for help with taxonomy, Jon Wilkins for constructive comments and clarifying discussions, and the "Scaling Group" at the University of New Mexico and the Santa Fe Institute for helpful discussions, especially Alex Herman.

# Figure captions

**Figure 1:** Plot of the logarithm of the ratio of total sleep time to total awake time per day, $\ln(t_S/t_A)$, versus the logarithm of body mass in Kg, $\ln(M)$. The slope computed using OLS regression is –0.16 (p<0.0001, n= 83, 95% CI: –0.21, -0.11). Note that the CI include –1/6 and –3/16 but exclude –1/4, supporting the hypothesis that cellular repair during sleep occurs primarily in the brain.

**Figure 2:** Plot of the logarithm of the ratio of total sleep time to total awake time per day, $\ln(t_S/t_A)$, versus the logarithm of brain mass in g, $\ln(M_b)$. The slope computed using OLS regression is –0.21 (p<0.0001, n=56, 95% CI: -0.28, -0.14). Note that the CI include –1/4, providing more evidence that sleep is driven primarily by the brain's metabolic rate.

**Figure 3:** Plot of the logarithm of sleep cycle time in minutes, the period between REM and non-REM sleep, versus the logarithm of body size in Kg, $\ln(M)$. The slope computed using OLS regression is 0.19 (p<0.0001, n=32, 95% CI: 0.14, 0.23). Note that the CI include 3/16 and 1/6 but exclude 1/4.

**Figure 4:** Plot of the logarithm of the ratio of REM sleep time to total sleep time per day, $\ln(R)$, versus the logarithm of body size, $\ln(M)$ in Kg. The slope computed using OLS regression is -6e-3 (p=0.74, n=61, 95% CI: -0.04, 0.03). The average value for $R$ is 0.17. Note that the CI for the slope are narrow and include zero. The large outlier on the right side of the plot corresponds to the pilot whale (*Globiocephalus scammoni*), a marine mammal.

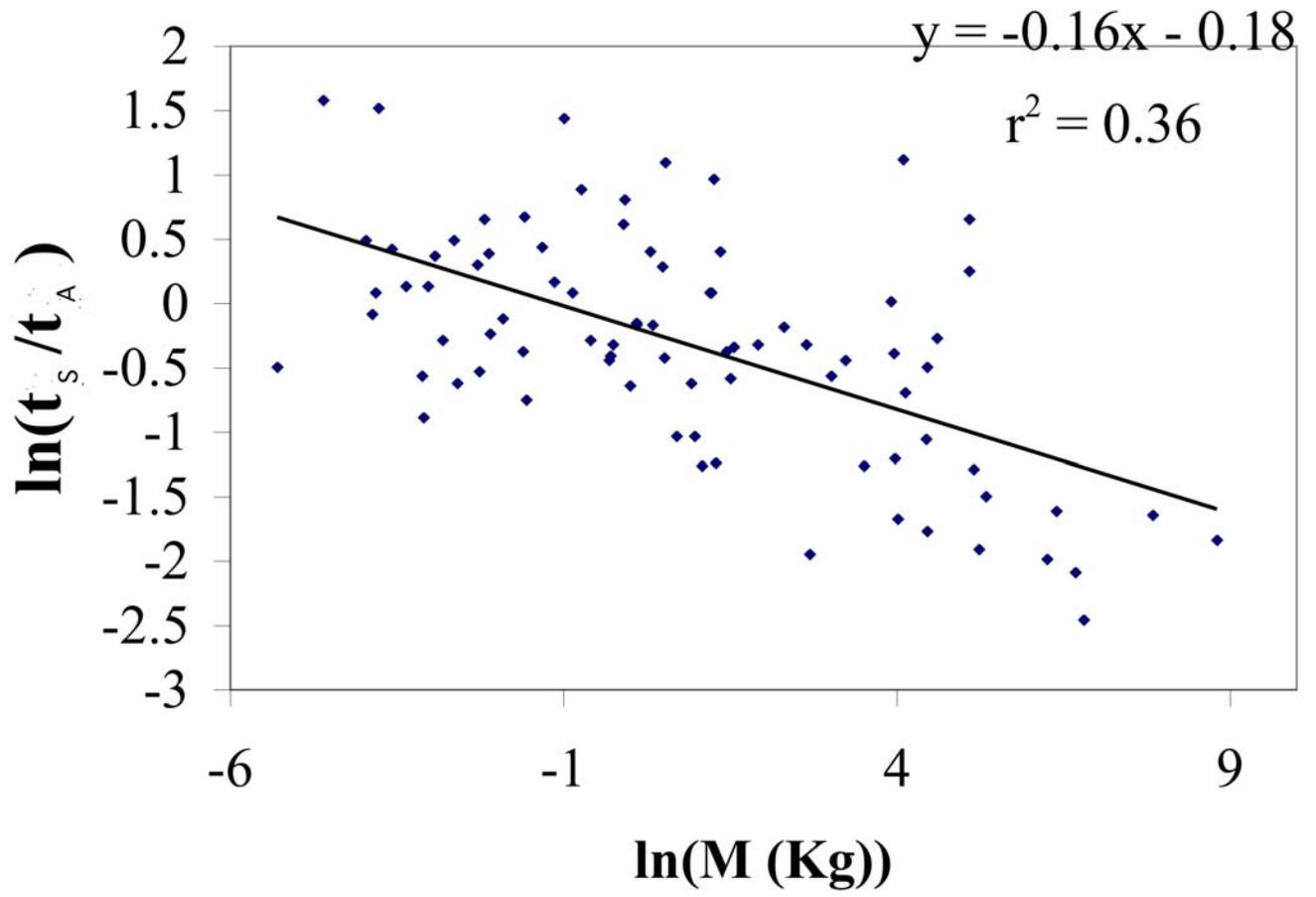

**Figure 1**

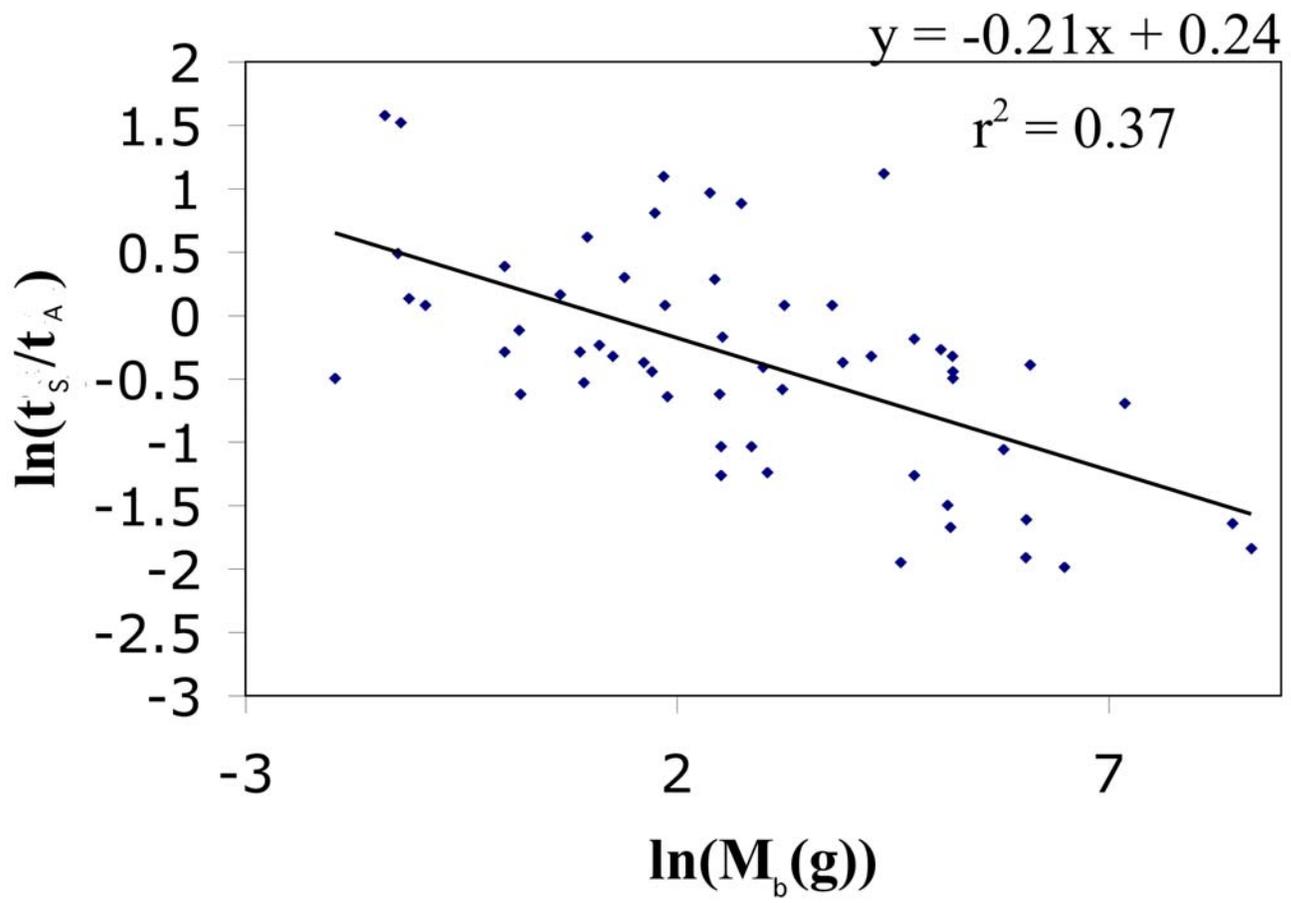

**Figure 2**

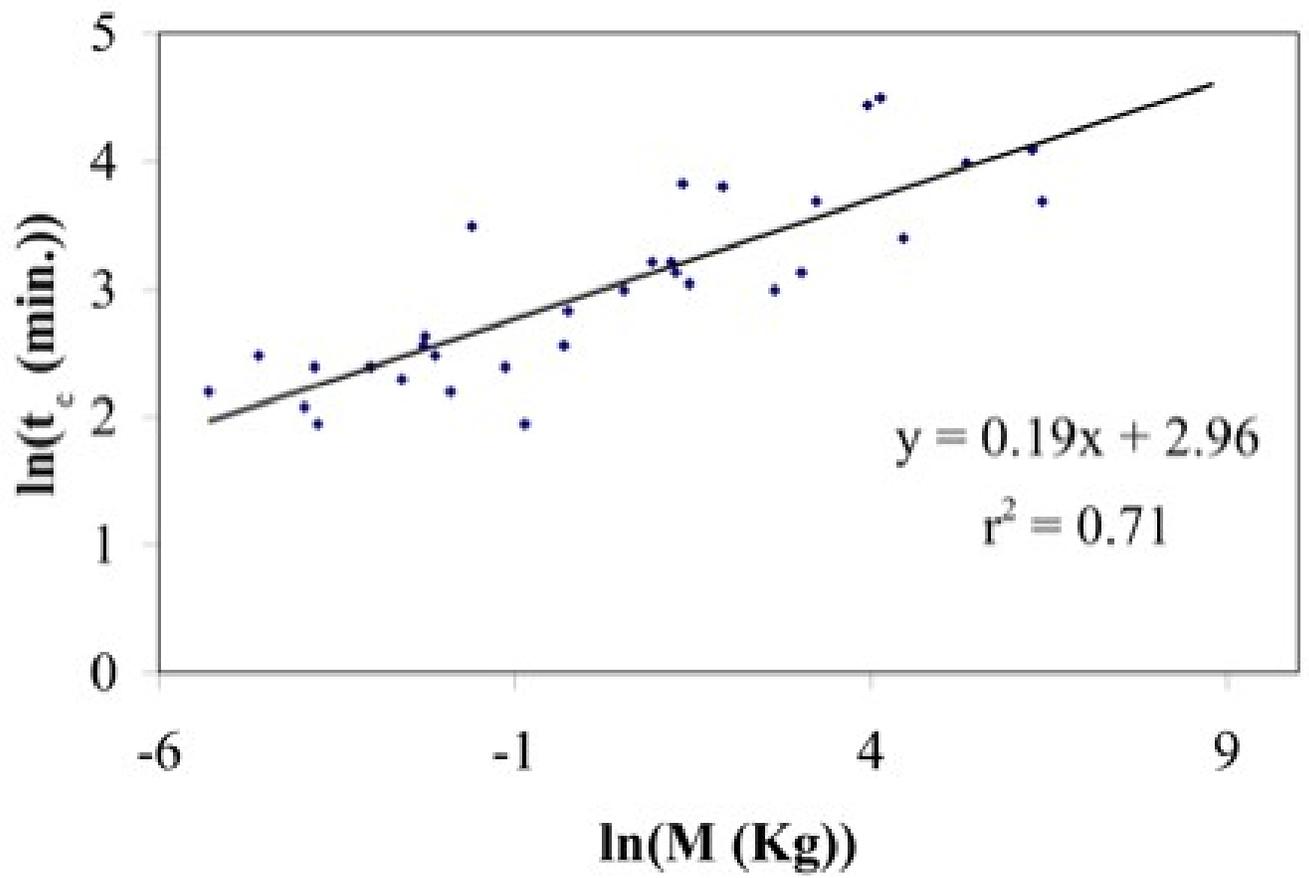

**Figure 3**

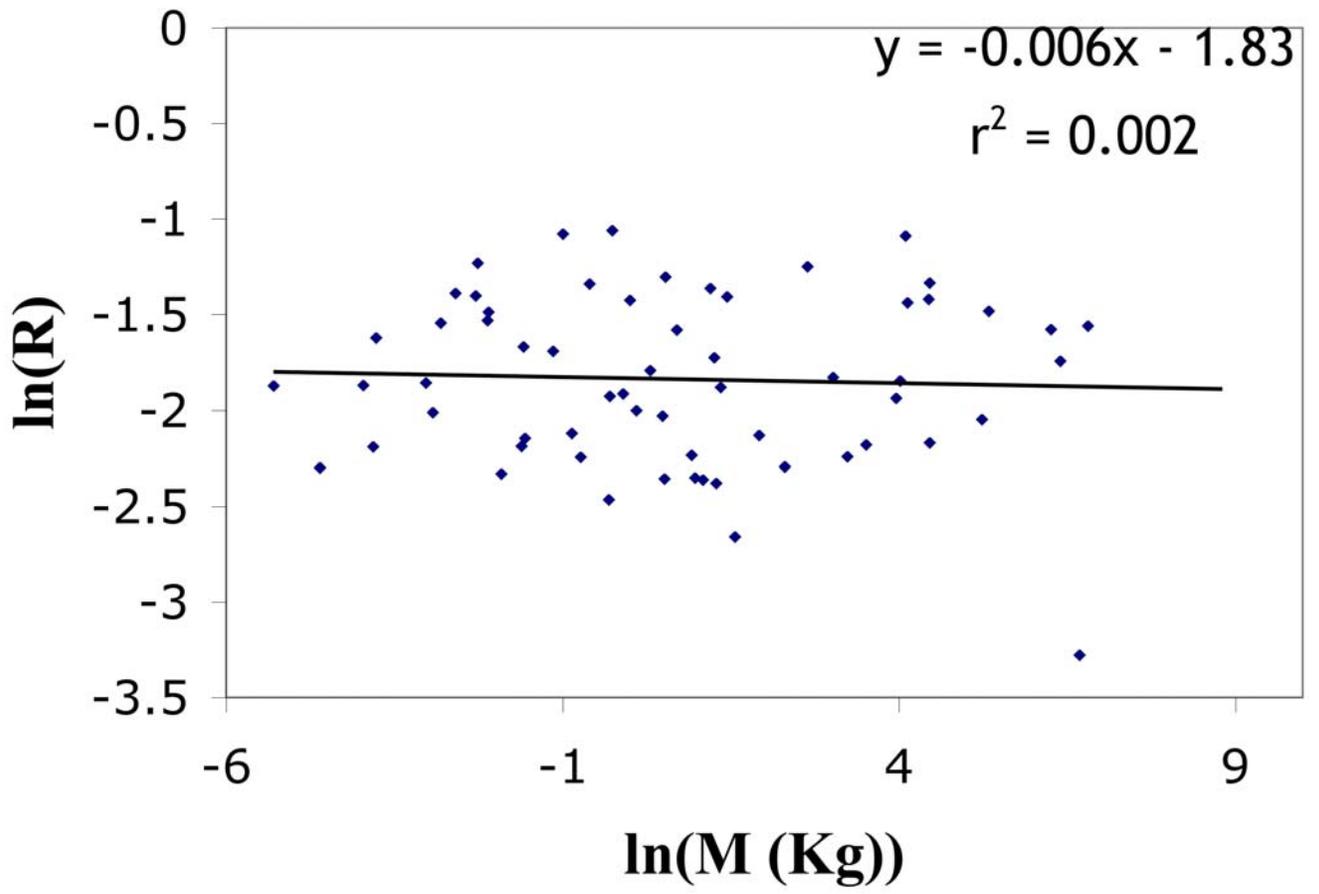

**Figure 4**